\documentclass[amsfonts,aps,preprint,nofootinbib]{revtex4-1}
\usepackage{hyperref}
\usepackage[dvips]{graphicx}
\usepackage[dvips]{epsfig}
\usepackage{latexsym}
\usepackage{amsmath,latexsym}
\usepackage[latin1]{inputenc}

\begin{document}
\title{String in AdS Black Hole: A Thermo Field Dynamic Approach }
\author{M. Botta Cantcheff}
\email{bottac@cern.ch, botta@fisica.unlp.edu.ar}
\affiliation{IFLP-CONICET CC 67, 1900,  La Plata, Buenos Aires, Argentina}
\affiliation{CERN, Theory Division, 1211 Geneva 23, Switzerland}
\author{Alexandre L. Gadelha}
\email{agadelha@ufba.br}
\affiliation{Instituto de F\'{\i}sica, Universidade Federal da Bahia,
Campus Universit\'ario de Ondina, CEP: 40210-340, Salvador, BA, Brasil}
\author{D\'afni F. Z. Marchioro}
\email{dafnimarchioro@unipampa.edu.br}
\author{Daniel Luiz Nedel}
\email{daniel.nedel@unipampa.edu.br}
\affiliation{Universidade Federal do Pampa, Campus Bag\'e, Travessa 45, n\'umero 1650 - Bairro Malafaia, CEP: 96413-170, Bag\'e, RS, Brasil}

\begin{abstract}
Based on Maldacena's description of an eternal AdS-black hole, we reassess the Thermo Field Dynamics (TFD) formalism in the context of the AdS/CFT correspondence. The model studied here involves the maximally extended AdS-Schwarschild solution and two (non-interacting) copies of the CFT associated to the global AdS spacetime, along with an extension of the string by imposing natural gluing conditions in the horizon.  We show that the gluing conditions in the horizon define a string boundary state which is identified with the TFD thermal vacuum, globally defined in the Kruskal extension of the AdS black hole. We emphasize the connection of this picture  with unitary SU(1,1) TFD formulation  and we show that information about the bulk and the conformal boundary is present in the SU(1,1) parameters. Using the unitary SU(1,1) TFD formulation, a canonical prescription for calculating the worldsheet real time thermal Green's function is made and the entropy associated with the entanglement of the two CFT's is calculated.
\end{abstract}

\preprint{CERN-PH-TH/2012-113}

\maketitle

\section{Introduction}

Since its advent, the holographic AdS/CFT correspondence has been exploited to study
the physics of non-Abelian quark-gluon plasmas at finite temperature from bulk gravitational
physics\cite{qg-plas1,qg-plas2,qg-plas3,qg-plas4}. More recently, it has been emphasized that
one can also exploit the gravitational description to understand the hydrodynamic regime of the
quark-gluon plasma and the quark Brownian motion in a conformal fluid \cite{deboer}. In these
applications, the string propagates in an asymptotically AdS space containing black holes, and
the quark physics is described by an open string stretching from the horizon up to a probe brane
placed to a short distance (related to the ultraviolet cut-off) from the conformal boundary. The
quark is naturally identified with the end of the string on the boundary, while near the Black Hole
horizon it is supposed to end on an effective membrane with thermal and dissipative properties, called stretched horizon. In this scenario, the  Hawking radiation induces random motion on the string end
point. Then, the  motion of the quark in the quark-gluon plasma is assumed to be described by a
Langevin equation, whose parameters can then be deducted by bulk calculations.
In principle, such investigations would require an exact first quantization of the string in an AdS black hole, which in general is a very difficult task. In spite of this problem, much progress has been made by studying small perturbations of the Nambu-Goto action up to quadratic order.\footnote{See \cite{thermalnoise}, \cite{giecold} and references therein.} This approximation corresponds to the non relativistic
limit and, in this context, the relativistic dynamics of the quark-gluon plasma at finite
temperature cannot be described.

In general, since the bulk geometry has an event horizon, the induced metric on
the string worldsheet also corresponds to a black hole geometry \cite{rindlerstretched}
and, owing to Hawking radiation, the problem of studying small fluctuations of the Nambu-Goto
action in AdS black holes reduces to the dynamics of two dimensional quantum field at  finite
temperature. So, in order to understand completely how the AdS/CFT at finite temperature works,
it is necessary to understand  this thermal field theory on the worldsheet.
In \cite{herzog} a prescription was formulated for computing the real-time Green's functions
at finite temperature. This was realized in the context of the Schwinger-Keldysh formalism,
by using the Kruskal extension of the AdS Schwarzschild spacetime.
In addition to Schwinger-Keldysh  formalism, there is another real time formalism
appropriated to explore the present context; it is the Thermo Field Dynamics (TFD)
formalism \cite{UmeTaka}. TFD provides a picture where the degrees of freedom behind
the horizon play an important role, which should be applied to describe strings in AdS
black holes. This point of view was noticed by Israel \cite{israel}, and put forth by
Maldacena in the holographic context \cite{eternal}. This is the picture that we intend
to study here.

TFD is a real time formalism conceived originally to deal with thermal systems
approaching them directly on their Fock space. The Takahashi and Umezawa original
proposal consists of a canonical quantum field theory that reproduces the
statistical averages of any system's observables \cite{UmeTaka}. In fact, from the
statistical mechanics point of view, considering an observable
operator $Q$, its statistical average is defined by the functional
$\omega(Q)=Tr(\rho Q)$ as
\begin{equation}
\left\langle Q \right\rangle = \frac{\omega(Q)}{\omega(1)}=\frac{Tr [Q\rho]}{Tr[\rho]}
\label{mes}
\end{equation}
where $\rho$ is the density operator of the system. Such functional is recognized
as a state in the algebraic statistical mechanics with the operators obeying a $C^{*}$
algebra. In this point of view, the algebra equipped with a functional admits a reducible
representation of the Hilbert space such as a Fock space
\cite{haag, emch, ojima81, landvan, sanka, skcc, torus, admB} and the statistical average
in TFD is written as
\begin{equation}
\left\langle Q \right\rangle = \frac{\omega(Q)}{\omega(1)}
=\left\langle 0(\beta)|Q |0(\beta)\right\rangle,
\label{mestfd}
\end{equation}
with $\left.|0(\beta)\right\rangle$ being the TFD thermal vacuum.
More generally, the central idea is to construct a quantum field theory whose vacuum contains
the information about the environment under which the system is subjected. Once observed that
temperature is introduced as an external parameter, it is verified that the thermal vacuum
appears as a boundary state in the doubled Fock space composed by the physical space of the
system and a copy of it. The expression defining such a state is called thermal state
condition \cite{UmeB} and it must contain all the information about the system to be
considered. Starting from a physical system at zero temperature, TFD's general procedure
consists of the doubling of degrees of freedom of the system and a suitable Bogoliubov
transformation to entangle such duplicated degrees of freedom. With the doubling, one obtains an
enlarged Hilbert space composed by the original and an auxiliary space, which is identical to the
original one and related to the so called tilde system. The enlarged Hilbert space is denoted by
a hat and is given by $\hat{\mathcal H} = {\mathcal H} \otimes \tilde{\mathcal H}$.
The original and tilde systems are related by a mapping called tilde conjugation rules
associated to the application of the Tomita-Takesaki modular operator of the statistical
mechanics algebraic approach \cite{ojima81, landvan}.

The Bogoliubov transformation is obtained using a generator in such a way that, in a
finite volume limit, the transformation is unitary and preserves the tilde conjugation rules.
The thermal effects arise from the vacuum correlation introduced by the transformation over
the enlarged system vacuum. This construction was the first one for TFD; however, one
can find a set of generators that maintains the thermal nature of the transformation. The
set of generators constructed to this end is shown to be a linear combination of operators that forms an oscillator
representation of SU(1,1) group for bosons and SU(2) for fermions \cite{UmeB, ChuUme}.
These sets for fermionic and bosonic systems can be combined in at least two different
ways, providing generalizations of the TFD approach. In one case the tilde conjugation
rules are preserved but the transformation, even in a finite volume limit, is non-unitary.
This construction was largely developed and its connection with other thermal field
theories is clear as one can see, for example, in Refs. \cite{UmeB} and \cite{Hen}.
In the second case the transformation is unitary in a finite volume limit, but the
tilde conjugation rules are not preserved. This is the so called general unitary
TFD formulation, and it is still under investigation.

Effectively, the general unitary formulation was applied to a physical
system in \cite{dbsu}, where it was perceived that a systematic study about the
formulation's implications was necessary. In references \cite{gusu11, gupr, Ksu11, gupos}
such analysis are carried out and the formulation was successfully applied to describe
superstrings at finite temperature\footnote{It is important to draw attention that TFD was also applied to string theory in
many others contexts in addition to those already mentioned \cite{Leb87,Leb881,Leb882,Leb89,FujNak891,FujNak891,FujNak892,FujNak893,
Leb901,Leb902,FujNak91,FujNak92,Fuj94,Fuj97,AGV00,IV00,AGV01,AGV031,
AGV032,AGN04,AGNPoS1, AGN052,GV,GMN,IV06,BGSV07,IV08,MN,NSV11,NV12}.}. Recently the
general unitary formulation was also applied in Refs \cite{brane-closed-pure, b-emer}, but considerations about interacting systems, dissipation, possible connection with other real time formalisms and with quantum statistical mechanics' algebraic approach appear as open
questions. Here we go forward towards a better understanding of these constructions.

In this work we show that the SU(1,1) TFD formulation arises naturally for strings propagating in an
AdS black hole geometry. This is achieved noting that, by gluing the string defined on the left quadrant
of the Kruskal diagram (with the string defined on the right one), we get a string boundary state,
which is exactly the thermal vacuum of TFD.  This is in fact an entangled state and in the context
of AdS/CFT, it is an entanglement of the two CFTs defined by the Kruskal extension of the AdS
black hole. We explore the general unitary SU(1,1) formulation of TFD to show
that all information of the system is present in the Bogoliubov parameters.
Also the TFD entropy operator is used to obtain the entropy of the system,
associated here to the presence of the horizon. Its high temperature limit is shown to be
in agreement with the Cardy formula.

Keeping in mind this whole framework and the motivations mentioned above,
we organize the present paper according to the following outline:
in Section II, we present the model (described originally in \cite{deboer}) and implement the Kruskal extension. In section III, in the spirit of reference \cite{thermalnoise} we present the
suitable generalization of the analytic continuation required to circumvent the conical singularity,
and then define a string in all spacetime by gluing up the strings of the left and right quadrant.
It is shown that the boundary state resulting from this procedure is an entangled state of elements
of the two CFTs. In Section IV, the general unitary SU(1,1) TFD formulation is presented, and it is
shown that the obtained vacuum state matches the boundary state coming from the gluing, which has
the information about the conditions imposed over the string defined in all spacetime. The entropy
is calculated and  it is interpreted as a result of the entanglement of the CFTs; also, the thermal
two-point functions are obtained. Finally, our final considerations and perspectives are presented
in section V.

\section{Open string in the AdS Black Hole}

In this section we describe the basic features
of the scenario that we are studying; a detailed discussion can be
found in ref \cite{deboer}. The model consists of putting a  probe
fundamental string stretched from the boundary to the horizon of the
following three dimensional asymptotically AdS black hole geometry:
\begin{equation}
ds^2 = -\frac{r^2 - r^{2}_{H}}{\ell^2} dt^2 +
\frac{r^2}{\ell^2} dX^2 + \frac{\ell^2}{r^2 - r^{2}_{H}} dr^2
\end{equation}
which is in fact the metric for a non rotating BTZ black hole. Here $t$
and $X$ are boundary coordinates. In this coordinate system, the open string
is suspended from the boundary at $r=\infty$ and straight down along the $r$
direction into the horizon at $r = r_H$.  Although we are focusing in this
simple three dimensional case, in \cite{deboer} the results are generalized
for $d$-dimensional spacetimes. The Hawking temperature is
\begin{equation}
T \equiv \frac{1}{\beta} = \frac{r_H}{2\pi \ell^2},
\end{equation}
where $\ell$ is the AdS radius.

Let us now write the string sigma model. For general planar $d$-dimensional
AdS black holes, the metric can always be written as follow
\begin{equation}
ds^2= g_{\mu\nu}dx^{\mu}dx^{\nu} + G_{IJ}dX^{I}dX^{J},
\end{equation}
where both  $g_{\mu\nu}$ and $G_{IJ}$ are independent of $X^I$ and $x^{\mu}=t,x$.
In the gauge that we are going to use, the worldsheet coordinates are
identified with the spacetime coordinates $t$ and $x$ and only the transversal
modes $X^I= X^I(t,x)$ are dynamical. In this gauge, the Nambu-Goto action can
be written as
\begin{equation}
S_{NG}= -\frac{1}{2\pi\alpha'}\int \sqrt{\det\gamma_{\mu\nu}},
\end{equation}
where the induced metric is $\gamma_{\mu\nu}= G_{IJ}\partial_{\mu}X^I\partial_{\nu}X^J$.
By expanding the Nambu-Goto action, we get a power series of $\partial_{t}X$ and
$\partial_{x}X$ which produces worldsheet interactions. As we are going to study only
small fluctuations of the equilibrium value $X^I=0$, only quadratic terms in the action
are considered. Since this approximation implies the regime $|\partial_{t}X^I| <<1 $
of small velocities, we are in fact  taking the non-relativistic limit. In the
quadratic approximation, the Nambu-Goto action can be written as
\begin{equation}
S_{NG} \approx -\frac{1}{4\pi\alpha'}\int\sqrt{g(x)}g^{\mu\nu}(x)G_{IJ}(x)\partial_{\mu}X^{I}\partial_{\nu}X^{J}
\label{ngquadratic}
\end{equation}
where $g(x)= \det g_{\mu\nu}$.

Following reference \cite{deboer}, one returns to $AdS_{3}$ case, where there is only
one transversal variable $X$, and writes the metric in terms of the tortoise coordinate $r_{\star}$,
\begin{eqnarray}
ds^2 = \frac{r^2 - r^{2}_{H}}{\ell^2} (-dt^2 + dr^{2}_{\star} )
+ \frac{r^2}{\ell^2} dX^2 ,
\nonumber
\\
r_{\star} \equiv \frac{\ell^2}{2r_H} \ln \left(\frac{r-r_H}{r+r_H} \right)
= \frac{\beta}{4\pi} \ln \left(\frac{r-r_H}{r+r_H} \right)
\end{eqnarray}
In this coordinate system, we can write the equations of motion in the quadratic approximation used in (\ref{ngquadratic}) as
\begin{equation}
\left[ -\partial^{2}_{t} + \frac{r^2 - r^{2}_{H}}{l^4 r^2} \partial_r (r^2 (r^2 - r^{2}_{H}) \partial_r ) \right] X(t,r) = 0
\label{eom}
\end{equation}
Defining dimensionless quantities
\begin{equation}
\rho \equiv \frac{r}{r_H}, \qquad \nu \equiv \frac{l^2 \omega}{r_H}
= \frac{\beta \omega}{2\pi},
\end{equation}
the linearly independent solutions to (\ref{eom}) are given by
\begin{equation}
X(t,r) = e^{-i\omega t} g_{\omega} (r) ,
\end{equation}
with
\begin{equation}
g^{(\pm)}_{\omega} =
\frac{1}{1 \pm i\nu} \frac{\rho \pm i\nu}{\rho}
\left(\frac{\rho -1}{\rho +1} \right)^{\pm i\nu/2}
= \frac{1}{1 \pm i\nu} \frac{\rho \pm i\nu}{\rho} e^{\pm i\omega r_{\star}}
\label{fomega}
\end{equation}

Because the string extends from the horizon of the black hole to the boundary,
a cut-off is needed to prevent UV divergences in the boundary. Also, we need
to regularize IR divergences at the horizon.

The UV divergences are controlled by imposing a Neumann boundary condition
$\partial_r X = 0$ at the cut-off surface:
\begin{equation}
\rho = \rho_c \gg 1 \qquad \mbox{or} \qquad r = r_c \equiv r_H \rho_c
\end{equation}
Solving for $\partial_{\rho} f_{\omega} |_{\rho = \rho_c} = 0$, we find the
value of $B$ in (\ref{sol}):
\begin{equation}
B =
\frac{1 - i\nu}{1 + i\nu} \frac{1 + i\rho_c \nu}{1 - i\rho_c \nu}
\left(\frac{\rho_c -1}{\rho_c +1}\right)^{i\nu} \equiv e^{i\theta_{\omega}}
\label{thetadef}
\end{equation}
which is a pure phase. Next it is necessary to regulate the IR divergence at
horizon ($\rho = 1$). This is achieved by  putting an IR cut-off at
$\rho_s= 1+2\epsilon$, with  $\epsilon <<1$. As argued in \cite{deboer}, the
effect of this regulator is to discretize the continuum spectrum which
naturally occurs when considering horizon dynamics.

After regularizing the theory, we can find a normalized basis of modes and
start quantizing $X(t,r)$ by expanding it in those modes:
\begin{equation}
X(t,r) = \sum_{\omega >0}
[a_{\omega} u_{\omega} (t, \rho)
+ a^{\dagger}_{\omega} [u_{\omega} (t,\rho)]^{\ast}]
\label{sol1}
\end{equation}
where
\begin{equation}
u_{\omega} (t,\rho ) =
\sqrt{\frac{\alpha' \beta}{2\ell^2 \omega \ln (1/\epsilon )}}
[g^{(+)}_{\omega} (\rho) + e^{i\theta_{\omega}} g^{(-)}_{\omega} (\rho )]
e^{-i\omega t}
\end{equation}
and the coefficients of the expansion $a_{\omega}$ satisfy the relations:
\begin{equation}
[a_{\omega} , a_{\omega '}] = [a^{\dagger}_{\omega} , a^{\dagger}_{\omega '}] = 0
\qquad
[a_{\omega} , a^{\dagger}_{\omega '} ] = \delta_{\omega \omega '}
\label{alg}
\end{equation}
Near the horizon ($\rho \sim 1$), the metric becomes plane and the
solution (\ref{fomega}) behaves like
\begin{equation}
g^{(\pm)}_{\omega} \sim e^{\pm i\omega r_{\star}}
\end{equation}
So, near the horizon the  solutions are written in terms of ingoing and
outgoing plane waves:
\begin{equation}
X_{R}(t,r) =
\sqrt{\frac{\alpha' \beta}{2\ell^2 \omega \ln (1/\epsilon )}} \sum_{\omega >0}
\{ a_{\omega} [e^{-i\omega (t- r_{\star})}
+ e^{i\theta_{\omega}} e^{-i\omega(t+ r_{\star} )}]\}.
\label{sol}
\end{equation}
Let us now further develop this model in order to show how the TFD structure
rises naturally in this scenario.
The solution (\ref{sol1}) can be written in terms of Kruskal coordinates,
where the structure of the full Penrose diagram for this AdS space becomes
apparent. Outside the horizon, there are two causally disconnected spacetime
geometries (both of which are asymptotically AdS), defining two CFTs.
In the Kruskal plane, the solutions are uniquely determined by their boundary
conditions at the two Minkowski boundaries, in the right ($R$) and, respectively,
left ($L$) quadrants of the Kruskal diagram. Also, in order to
define the string in all spacetime, it is necessary to perform an appropriate
analytical continuation  in the Kruskal variables $U$ and $V$. The Kruskal
coordinates are defined in terms of $t$ and $r_{\star}$ by the transformation
\begin{eqnarray}
t = \ln \left(\frac{V}{-U} \right) \qquad r_{\star} = \ln (V(-U))
\label{kruskal}
\end{eqnarray}
where $(U, V)$ are defined such that  ($V >$ 0, $U < 0$) in the right quadrant
and ($V < 0$, $U > 0$) in the left quadrant.

Note that, as any function of $t$ and $r$ can be written in terms of $U$ and
$V$, we don't need to write the metric in terms of $U$ and $V$ and solve again
the equation of motion. We can solve the string equation of motion separately
in the $R$ and $L$ quadrants and obtain one set of mode functions in each quadrant.
So, if we define the solution (\ref{sol1}) in the $R$ quadrant, we just need to write
the respective solution in the $L$ quadrant defining it as a copy of the $R$ solution:
\begin{equation}
X_{L}(\tilde{t},r) = \sum_{\omega >0}
\{ \tilde{a}_{\omega} \tilde{u}_{\omega} (\tilde{t}, \rho)
+ \tilde{a}^{\dagger}_{\omega} [\tilde{u}_{\omega} (\tilde{t}, \rho)]^{\ast} \}
\label{soll}
\end{equation}
where
\begin{equation}
\tilde{u}_{\omega} (\tilde{t},\rho ) =
\sqrt{\frac{\alpha' \beta}{2 \ell^{2} \omega \ln (1/\epsilon )}}
[g^{(+)}_{\omega} (\rho) + e^{i\tilde{\theta}_{\omega}} g^{(-)}_{\omega} (\rho )]
e^{i\omega \tilde{t}}
\end{equation}

In order to relate  these solutions, we need to impose boundary conditions at the
horizon $U=V=0$, where the transformation (\ref {kruskal}) is singular. This is the
topic of the next section. By doing the Kruskal extension, we have defined two CFTs.
Although classically these CFTs are causally disconnected, we are going to show
in the next sections that states of a CFT defined in one quadrant play a particular
role in the expectation values of observables defined in the other quadrant.

\section{Building up the extended string from gluing}

The Kruskal extension defines two  asymptotically AdS spaces and
two causally disconnected CFTs.  By defining the string in all spacetime,
we can quantum connect these two CFTs. In particular, we are going to show
now that the string vacuum is an entangled state of states of the
two CFTs and it is the thermal vacuum of the TFD.

In order to define continuously the string in all spacetime by
connecting the solution in the $L$ quadrant to that in the R quadrant, we need
to avoid the singularity at $U = V = 0$ by performing a well known analytic
continuation. In terms of the old coordinates, this analytical continuation relates
$\tilde{t} - t= i\frac{\beta}{2}$, where $\tilde{t}$ is the time coordinate
of the $L$ quadrant of the spacetime. The resulting thermal field theory
ultimately reproduces the contour correlation functions for
a specific choice of the Schwinger-Keldysh contour. As pointed out in \cite{thermalnoise},
this analytical continuation can be generalized in order to take into account deformations
of the usual Schwinger-Keldysh contour. Inspired by that observation, but taking in mind
the TFD approach, our proposal is considering $\tilde{t} - t= i\alpha\beta$, for $\alpha$
a complex parameter. As it will be clear in the next section, the unitarity of the SU(1,1)
TFD formulation will imply in $\alpha+ \alpha^* =1$. Also, this constraint ensures the
Kubo-Martin-Schwinger (KMS) condition and therefore the usual time periodicity of the
correlation functions.

Near the horizon ($\rho \sim 1$), the solutions (\ref{sol1}) and
(\ref{soll}) simplify to
\begin{eqnarray}
X_{R}(t,r) =
\sqrt{\frac{\alpha' \beta}{2\ell^2 \omega \ln (1/\epsilon )}}\sum_{\omega > 0}
\left\{ a_{\omega} \left[e^{-i\omega (t- r_{\star})}
+  e^{i\theta_{\omega}} e^{-i\omega(t+ r_{\star} )}\right]
\right.
\nonumber
\\
\left.
+ a^{\dagger}_{\omega} \left[ e^{i\omega (t - r_{\star})}
+ e^{-i\theta_{\omega}}  e^{i\omega(t + r_{\star})}\right] \right\},
\label{xr}
\end{eqnarray}
\begin{eqnarray}
X_{L}(t,r) =
\sqrt{\frac{\alpha' \beta}{2\ell^2 \omega \ln (1/\epsilon )}}\sum_{\omega >0}
\left\{ \tilde{a}_{\omega} \left[ e^{i\omega (\tilde{t} - r_{\star} )}
+ e^{-i\tilde{\theta}_{\omega}}  e^{i\omega(\tilde{t}+r_{\star})}\right]
\right.
\nonumber
\\
\left.
+ \tilde{a}^{\dagger}_{\omega} \left[e^{-i\omega (\tilde{t}-r_{\star})} + e^{i\tilde{\theta}_{\omega}}  e^{-i\omega (\tilde{t}+r_{\star})}\right] \right\}.
\label{xl}
\end{eqnarray}
Note that owing to reverse clockwise direction of the L quadrant,
while the $a$ operator annihilates outgoing modes at the horizon in $R$ quadrant,
the $a$ operator annihilates ingoing modes in the L quadrant.
We are going to impose gluing conditions on a state called $|B_{hor} \rangle$,
which represents that the string in the $R$ quadrant is connected with its dual
copy in the $L$ quadrant. These conditions are
\begin{eqnarray}
\left(X_L (t, r) |_{\rho =1} - X_R (t,r)|_{\rho=1} \right) | B_{hor} \rangle = 0
\label{glcond}
\end{eqnarray}
We avoid the singularity at  $\rho =1$ using Kruskal coordinates and performing
the analytic continuation $\tilde{t}= t+ i\alpha\beta$. Solving for the gluing
conditions, we have
\begin{eqnarray}
\left(\tilde{a}_{\omega}e^{-\omega\alpha\beta}
- a^{\dagger}_{\omega} \right) | B_{hor} \rangle = 0
\nonumber
\\
e^{-i\theta_{\omega}} \left( \tilde{a}_{\omega}e^{-\omega\alpha\beta}
- a^{\dagger}_{\omega} \right) |B_{hor} \rangle = 0
\nonumber
\\
\left( a_{\omega} -
\tilde{a}^{\dagger}_{\omega} e^{\omega\alpha\beta}  \right)
|B_{hor} \rangle = 0
\nonumber
\\
e^{i\theta_{\omega}}
\left( a_{\omega}- \tilde{a}^{\dagger}_{\omega}e^{\omega\alpha\beta}\right)
|B_{hor} \rangle = 0
\label{glconds}
\end{eqnarray}

The state $|B_{hor} \rangle$, which satisfy the equations (\ref{ts}), is:
\begin{equation}
|B_{hor} \rangle =
N \exp \left[ \sum_{\omega} e^{-\alpha\beta\omega} a^{\dagger}_{\omega} \tilde{a}^{\dagger}_{\omega} \right] |0, \tilde{0} \rangle, \label{ts}
\end{equation}
where $N$ is a normalization factor.

The boundary sate $|B_{hor} \rangle$ is a string entangled state, which entangles
string states defined in the L and R quadrants. In the AdS/CFT context,
this state entangles the two boundary CFTs, in particular the heavy quark states
defined by the endpoint of the string at the two asymptotic boundaries.

Notice that these states are built up by imposing boundary conditions,
referred here as gluing conditions. A similar construction in the Minkowski
spacetime was done in Ref. \cite{braneopen}. It has been also argued that
these states are equivalent to ordinary boundary states in the \emph{closed string}
Hilbert space under a worldsheet transformation of the one-loop diagram in Euclidean
time \cite{braneopen}.

In the next section we are going to show that $|B_{hor} \rangle$ can also
be  interpreted as the thermal vacuum in the point of view of Thermo Field
Dynamics (TFD), where the equilibrium temperature is the Hawking temperature.
Expectation values of the R string states on this state correspond to statistical
averages in an ensemble of open string states. In the context of the AdS/CFT conjecture,
this imply that we  can calculate non-perturbatively thermodynamical properties of the
CFT through the calculation of the expectation values in this state.

\section{The Bogoliubov Transformation, Entropy Operator and Thermal Two Point Function}

Let us show now that the boundary state defined in the previous section can be achieved
from a Bogoliubov transformation and  that this state is also defined by
an entanglement entropy operator. Indeed, the TFD approach appears
naturally in this scenario. As a matter of fact, the general unitary $SU(1,1)$ TFD
formulation will be used in order to fit the generalization proposed for the analytic
continuation.

The presence of the horizon defines two string Hilbert spaces,
related to the two regions $L$ and $R$. In the following we will
refer to the elements of these regions as tilde and non-tilde, respectively.
The total Hilbert space is the tensor product of the two spaces
${\cal H}_{o}\otimes\widetilde{{\cal H}}_{o}$, where in this case
${\cal H} _{o}$ is the Hilbert space built with cyclic applications of the
operators $a_{\omega}, a^{\dagger}_\omega$ while the $\widetilde{{\cal H}}_{o}$
Hilbert space is related to the $\tilde{a}_{\omega}, \tilde{a}^{\dagger}_\omega$
operators. The standard vacuum in this extended theory is defined by
\begin{equation}
{a}_{\omega}\left.\left|0\right\rangle\!\right\rangle =
\tilde{a}_{\omega}\left.\left|0\right\rangle\!\right\rangle = 0,
\end{equation}
with
$\left.\left|0\right\rangle\!\right\rangle =|0\rangle \otimes|\tilde{0} \rangle$
as usual.

Owing to the reverse clockwise direction of the L quadrant, the total worldsheet
Hamiltonian is defined as $\widehat{H}= H-\widetilde{ H}$, where the $H$ Hamiltonian
is proportional to number operator
$N_{\omega}= a_{\omega}^{\dagger} a_{\omega}$
and $\widetilde{ H}$ is proportional to
$\widetilde{N}_{\omega}=\tilde{a}_{\omega}^{\dagger}\tilde{a}_{\omega}$.
The transformation generator that will be considered here is
\cite{ChuUme, UmeB, dbsu}\footnote{It must be noticed that the generator
used here corresponds to the one that generates the inverse Bogoliubov
transformation in \cite{dbsu} and \cite{gusu11}, for example.}.
\begin{equation}
{\bf G}\left( \theta \right) =\sum_{i=1}^{3} {\bf g}_{i}\left( \theta \right) ,
\label{G}
\end{equation}
with,
\begin{eqnarray}
{\bf g}_{1}\left( \theta \right) &=&
-\sum_{\omega=1}\theta _{1_{\omega}}\left( a_{\omega}\tilde{a}_{\omega}
+\tilde{a}_{\omega}^{\dagger }a^{\dagger }_{\omega}\right) ,
\\
{\bf g}_{2}\left( \theta \right) &=&
-\sum_{\omega=1}i\theta_{2_{\omega}}\left( a_{\omega}\tilde{a}_{\omega}
-\tilde{a}^{\dagger }_{\omega}a^{\dagger }_{\omega}\right) ,
\\
{\bf g}_{3}\left( \theta \right) &=&
-\sum_{\omega=1}\theta _{3_{\omega}}\left( a^{\dagger }_{\omega}a_{\omega}
+\tilde{a}_{\omega}\tilde{a}^{\dagger }\right).
\end{eqnarray}
Here $\theta_{j_{\omega}}$, $j=1,2,3$ denotes the set of transformation parameters.
The generators written in the last equations satisfy an su(1,1) algebra, and (\ref{G})
can be rearranged as
\begin{equation}
{\bf G}(\gamma) =- \sum_{\omega=1}
\left[ \gamma _{1_{\omega}}\tilde{a}_{\omega}^{\dagger }a^{\dagger }_{\omega}
- \gamma_{2_{\omega}} a_{\omega}\cdot \tilde{a}_{\omega}
+ \gamma_{3_{\omega}}\left( a^{\dagger }_{\omega} a_{\omega}
+ \tilde{a}_{\omega}\tilde{a}^{\dagger }_{\omega}\right)\right] ,
\label{ge}
\end{equation}
where the $\gamma$'s coefficients are defined as
\begin{equation}
\gamma _{1_{\omega}} =\theta _{1_{\omega}}-i\theta
_{2_{\omega}}, \qquad \gamma_{2_{\omega}} =-\gamma
_{1_{\omega}}^{*},\qquad \gamma_{3_{\omega}} =\theta
_{3_{\omega}}.
\label{gammadef}
\end{equation}
This generator carries out a unitary and canonical transformation, such that
the creation and annihilation operators transform according to \cite{gusu11}
\begin{eqnarray}
\left(
\begin{array}{c}
a_{\omega}(\gamma) \\
\breve{a}^{\dagger}_{\omega}(\gamma)
\end{array}
\right) &=&e^{-i{\bf G}}\left(
\begin{array}{c}
a_{\omega} \\
\tilde{a}^{\dagger }_{\omega}
\end{array}
\right) e^{i{\bf G}}={\mathbb B}_{\omega}\left(
\begin{array}{c}
a_{\omega} \\
\tilde{a}^{\dagger }_{\omega}
\end{array}
\right) ,
\nonumber
\\
\left(
\begin{array}{cc}
a^{\dagger}_{\omega}(\gamma) & -\breve{a}_{\omega}(\gamma)
\end{array}
\right) &=&\left(
\begin{array}{cc}
a^{\dagger }_{\omega} & -\tilde{a}_{\omega}
\end{array}
\right) {\mathbb B}^{-1}_{\omega},
\label{tbti}
\end{eqnarray}
where the $SU(1,1)$ matrix transformation is given by
\begin{eqnarray}
{\mathbb B}_{\omega}=\left(
\begin{array}{cc}
\mathfrak{u}_{\omega} & \mathfrak{v}_{\omega} \\
\mathfrak{v}^{*}_{\omega} & \mathfrak{u}^{*}_{\omega}
\end{array}
\right) ,
\qquad |\mathfrak{u}_{\omega}|^{2} - | \mathfrak{v}_{\omega}|^{2}=1,
\label{tbm}
\end{eqnarray}
with elements \cite{dbsu}
\begin{equation}
\mathfrak{u}_{\omega}=\cosh \left( i\Gamma_{\omega} \right) +\frac{\gamma
_{3_{\omega}}}{\Gamma_{\omega} } \sinh\left(i\Gamma_{\omega} \right),
\qquad
\mathfrak{v}_{\omega}=-\frac{\gamma _{1_{\omega}}}{\Gamma_{\omega} }\sinh\left(
i\Gamma_{\omega} \right) , \label{uvexp}
\end{equation}
and $\Gamma_{\omega}$ is defined by the following relation
\begin{equation}
\Gamma ^{2}_{\omega}=\gamma _{1_{\omega}}\gamma _{2_{\omega}}+\gamma
_{3_{\omega}}^{2}.
\label{Gadef}
\end{equation}
A quite convenient way to write the Bogoliubov transformation matrix (\ref{tbm})
arises if we make the polar decomposition
$\mathfrak{u}_{\omega}=|\mathfrak{u}_{\omega}|e^{i\varphi_{\omega}}$,
$\mathfrak{v}_{\omega}=|\mathfrak{v}_{\omega}|e^{i\phi_{\omega}}$,
and rewrite the matrix elements in terms of the new parameters
\begin{equation}
f_{\omega}=\frac{|\mathfrak{v}_{\omega}|^{2}}{|\mathfrak{u}_{\omega}|^{2}},
\quad
\alpha_{\omega}=
\frac{\log(\frac{\mathfrak{v}_{\omega}}{\mathfrak{u}_{\omega}})}{\log(f_{\omega})}
= \frac{1}{2}+i\frac{(\phi_{\omega}-\varphi_{\omega})}{\log(f_{\omega})},
\quad
s_{\omega}=i\varphi_{\omega}=
\frac{1}{2}\log\left(\frac{\mathfrak{u}_{\omega}}{\mathfrak{u}_{\omega}^{*}}\right).
\label{sfap}
\end{equation}
In fact, with these steps we can present the Bogoliubov matrix as \cite{Hen, UmeB}
\begin{equation}
{\mathbb B}_{n}=\frac{1}{\sqrt{1-f_{\omega}}}\left(
\begin{array}{cc}
e^{s_{\omega}} & -f_{\omega}^{\alpha_{\omega}}e^{s_{\omega}} \\
-f_{\omega}^{\alpha_{\omega}^{*}}e^{-s_{\omega}} & e^{-s_{\omega}}
\end{array}
\right),
\label{sfapM}
\end{equation}
with $\alpha_{\omega} + \alpha^{*}_{\omega} = 1$.

As the Bogoliubov transformation is canonical, the gamma dependent
operators obey the same commutation relations of $(\ref{alg})$. One can define a
vacuum state for the transformed system as the state satisfying
\begin{eqnarray}
a_{\omega}(\gamma)\left |0(\gamma)\right\rangle &=& \widetilde
{a}_{\omega}(\gamma)\left |0(\gamma)\right\rangle = 0.
\end{eqnarray}
This expression, together with $(\ref{sfap})$ and $(\ref{sfapM})$, gives rise to
the following conditions:
\begin{eqnarray}
e^{s_{\omega}}\left[a_{\omega}
-f_{\omega}^{\alpha_{\omega}}\tilde{a}^{\dagger}_\omega\right]
\left|0(\gamma)\right\rangle
&=&0,
\label{cond1}
\\
e^{-s_{\omega}}\left[\tilde{a}_{\omega}
-f_{\omega}^{\alpha_{\omega}^{*}}{a}^{\dagger}_{\omega}\right]
\left|0(\gamma)\right\rangle
&=&0,
\label{cond2}
\end{eqnarray}
At this moment, it is already possible to compare the conditions presented above
with those defining the state $|B_{hor} \rangle$ in $(\ref{glconds})$.
However, let us explore a bit more the TFD formalism used here.
The inverse of the Bogoliubov transformations $(\ref{tbti})$ allow one to obtain
that the number of modes of the string defined in the R quadrant is
\begin{equation}
\bar{N}_{\omega}(\theta)=\left\langle 0(\beta)|a_{\omega} a_{\omega}^{\dagger}|0(\theta)\right\rangle
=|\mathfrak{v}_{\omega}|^{2}=\frac{\gamma _{1_{\omega}}\gamma_{2_{\omega}}}
{\Gamma^2_{\omega}}\sinh^{2}(i\Gamma_{\omega}) = \frac{f_{\omega}}{1-f_{\omega}}
\label{nu}
\end{equation}
and similarly for the modes of type $\tilde{a}_n$ in the L quadrant.
The transformation also entangles the states of the two independent Hilbert spaces
\cite{Vitent,Adement}, and gives us a structure to the new vacuum,
$\left|0(\gamma)\right\rangle$, as follows
\begin{equation}
\left |0(\gamma)\right\rangle =
e^{i{{\bf G}}} \left.\left|0\right\rangle \! \right\rangle =
\prod_{\omega=1}\left[e^{-s_{\omega}}\sqrt{1-f_{\omega}}\;
e^{f_{\omega}^{\alpha_{\omega}}\, a_{\omega}^{\dagger}{\tilde a}_{\omega}^{\dagger}} \right]
\left.\left|0\right\rangle\!\right\rangle.
\label{tva}
\end{equation}
For a suitable parameter choice, the state (\ref{tva}) is the string state
defined in (\ref{ts}), and the relations (\ref{cond1}), (\ref{cond2}) are the
relations (\ref{glconds}), called thermal state conditions in the TFD formulation.
Indeed, transformation parameters, as expected, encode information about
the environment under which the system is subject (in our study, information
about the model presented). Furthermore, TFD general approaches usually consider the
$s_{\omega}$ and $\alpha_{\omega}$ parameters as being the same for all modes and
deal with them as free parameters that can be fixed suitably for each situation.
For the application considered here, it will be possible to verify that the
$s_{\omega}$ parameter is no longer free, once it seems to contain information
about the CFT boundary conditions. Note that $s_{\omega}$ is related to $\theta_{\omega}$,
which defines the Neumann boundary conditions at cut off surface $\rho=\rho_c$.
The $\alpha$ parameter is the only one that can be considered free in some sense.
Rather, it is possible to consider it as being the same for all string modes
($\alpha_{\omega} \rightarrow \alpha$) but also constrained by the relation
$\alpha + \alpha^{*} = 1$, that comes directly from the general TFD construction.
From the thermo field point of view, the $\alpha$ constraint guarantees the
KMS conditions as it can be verified directly following the proof for the
non-unitary TFD formulation in \cite{UmeB},
which might be useful in order to simplify the treatment of more engaged situations
such as those where interactions or non-equilibrium effects are explicitly
considered, as it is the case of the parameters' choice of the non-unitary
TFD formulation.\footnote{In fact, the non-unitary TFD formulation and that presented here
share many formal features as it was pointed out in references \cite{ChuUme}
and \cite{gupr} for instance.}
On the other hand, from the geometry's perspective, this constraint ensures
the  appropriate time periodicity, which is necessary to circumvent the conical
singularity. Furthermore, $\alpha$ is related to the trace cyclicity
in thermal statistical averages, as it seems to be the case here.\footnote{Notice that
in expression (\ref{mes}), $Tr [Q\rho ] = Tr [\rho^{(1-\alpha)}Q\rho^{(\alpha)}]$.}
However, as it will be shown, the expectation value of the system's observables at thermal
equilibrium does not depend on $\alpha$. Finally, the $f_{\omega}$ parameter
will contain information about the thermal distribution of the string modes, as it will
be noticed.

Before specifying the parameters, let us introduce a very interesting
operator which arises in this formalism. Consider the following gamma
dependent operators defined on the R and L sectors, respectively
\begin{eqnarray}
K&=&-\sum_{\omega=1}\left[ a_{\omega}^{\dagger }a_{\omega}
\ln\left(\frac{\gamma _{1_{\omega}}\gamma _{2_{\omega}}}
{\Gamma_{\omega}^{2}}\sinh^{2}\left( i\Gamma _{\omega}\right)\right)
-a_{\omega}a_{\omega}^{\dagger }
\ln \left( 1+ \frac{\gamma _{1_{\omega}}\gamma_{2_{\omega}}}
{\Gamma_{\omega}^{2}}\sinh^{2}\left( i\Gamma_{\omega}\right)\right)\right],
\label{K}
\\
\widetilde{K}&=&-\sum_{\omega=1}\left[ \tilde{a}_{\omega}^{\dagger }\tilde{a}_{\omega}
\ln\left(\frac{\gamma _{1_{\omega}}\gamma _{2_{\omega}}}
{\Gamma_{\omega}^{2}}\sinh^{2}\left( i\Gamma _{\omega}\right)\right)
-\tilde{a}_{\omega}\tilde{a}_{\omega}^{\dagger }
\ln \left( 1+ \frac{\gamma _{1_{\omega}}\gamma_{2_{\omega}}}
{\Gamma_{\omega}^{2}}\sinh^{2}\left( i\Gamma_{\omega}\right)\right)\right],
\label{Ktil}
\end{eqnarray}
These operators are the entropy operators for the general unitary TFD
formulation \cite{Ksu11}. The extended operator
$\widehat{K}= K - \widetilde{K}$
commutes with the Bogoliubov transformation generator (\ref{ge}).

The expectation value of the operator $K$ evaluated at the gamma dependent vacuum state
can be calculated by usual methods and the result is given below:
\begin{equation}
S(\gamma) = \left\langle 0(\theta)\right| K \left|0(\theta)\right\rangle
=\sum_{\omega=1}\left\{(1+\bar{N}_{\omega})\ln(1+\bar{N}_{\omega})
- \bar{N}_{\omega}\ln\bar{N}_{\omega}\right\},
\label{Kgamma}
\end{equation}
where $\bar{N}_{\omega}$ was defined by (\ref{nu}); a similar expression is
obtained for the L quadrant. The vacuum state (\ref{tva}) can be rewritten
using these operators as follows\footnote{Effectively, the
state obtained using the $K$ operator differs from that arising from the use
of the Bogoliubov generator by a phase, as it was pointed out in
\cite{Ksu11}.}
\begin{equation}
\left|0(\gamma)\right\rangle =
e^{\alpha K}e^{\sum_{\omega=1}a^{\dagger}_{\omega}\tilde{a}^{\dagger}_{\omega}}
\left.\left|0\right\rangle\!\right\rangle
= e^{\alpha\widetilde{K}}e^{\sum_{\omega=1}a^{\dagger}_{\omega}\tilde{a}^{\dagger}_{\omega}}
\left.\left|0\right\rangle\!\right\rangle. \label{entropiavacum}
\end{equation}
or
\begin{equation}
\left|0(\gamma)\right\rangle =
\sum_{\omega=1}{W_{\omega}}^{\alpha}\left|\omega,\tilde{\omega}\right\rangle,
\label{gammavacW}
\end{equation}
with
\begin{equation}
W_{\omega}=
\prod_{n=0}
\frac{\left(|\mathfrak{v}_{\omega}|^{2}\right)^{n_{\omega}}}
{\left(|\mathfrak{u}_{\omega}|^{2}\right)^{n_{\omega}+1}}
=(1-f_{\omega})\prod_{n=0} (f_{\omega})^{n_{\omega}},
\qquad
\sum_{\omega=1}W_{\omega}=1,
\end{equation}
and (\ref{Kgamma}) assumes the following form
\begin{equation}
S(\gamma) = -\sum_{\omega=1} W_{\omega} \ln W_{\omega}
\end{equation}
As those results can be obtained by using $\tilde{K}$, we have the entropy
operator defined in both quadrants.
The whole vacuum $|0(\gamma)\rangle$ is an entanglement of states from
both quadrants as it can be noticed explicitly in (\ref{gammavacW}), and therefore
expectation value of $K$ furnishes the entanglement entropy of the system.

The expressions presented earlier show that the presence of a horizon produces string
entanglement entropy and the origin of the entanglement is the environment,
in contrast with the usual quantum mechanical point of view, which attributes to the
environment the loss of the entanglement. In this picture, the origin of
macroscopic dissipation in the CFT side is related to the open string entanglement
entropy caused by the AdS geometry.

Finally, the vacuum can be completely defined by minimizing the free energy
\begin{equation}
\mathcal{F} = \mathcal{U} - \frac{1}{\beta }\mathcal{S}
\label{f}
\end{equation}
with respect to the transformation parameters \cite{UmeTaka},
defining the open string thermal vacuum. Here $\mathcal{U}$ is given by the
vacuum expectation value of the open string Hamiltonian in the thermal vacuum and
$\mathcal{S}$ is the entropy given in (\ref{Kgamma}).
The solution for the Bogoliubov transformation parameters is
given by the Bose-Einstein distribution. In fact, in this context
$f_{\omega} = e^{-\beta\omega}$, and the expression (\ref{nu}) is
now given by
\begin{equation}
\bar{N}_{\omega}(\beta) = \frac{e^{-\beta\omega}}{1-e^{-\beta\omega}}.
\label{bose}
\end{equation}

Taking into account these solutions and further relating the $s_{\omega}$ parameter
in (\ref{sfap}) with $\theta_{\omega}$ defined in (\ref{thetadef}),
the state $(\ref{tva})$ is written as
\begin{equation}
\left |0(\beta)\right\rangle =
\prod_{n=1}e^{-i\theta_{n} }\sqrt{1-e^{-\beta n}}
\exp\left[{\sum_{\omega=1}
e^{-\alpha\omega\beta}\, a_{\omega}^{\dagger}{\tilde a}_{\omega}^{\dagger}}\right]
\left.\left|0\right\rangle\!\right\rangle,
\label{tsvaccum}
\end{equation}
(compare with that proposed in (\ref{ts}) as a solution of the gluing condition),
and it is in fact the open string thermal vacuum for the system under consideration.
Even knowing that the parameters $\theta_{\omega}$ are phases and $\alpha$ does not
interfere in the thermal physics, the TFD general formulation shows itself quite
fancy since it furnishes naturally a state with all the information of the model.

Also notice that once the Bogoliubov transformation parameter has been fixed as above,
the expression (\ref{Kgamma}) for the entropy takes the following form:
\begin{equation}
\mathcal{S}(\beta) = \sum_{\omega=1}\beta\omega\bar{N}_{\omega} +\ln Z,
\label{Ss}
\end{equation}
with
\begin{equation}
Z= \prod_{\omega=1}\frac{1}{1-e^{-\beta\omega}}.
\label{Z}
\end{equation}
Finite temperature supposedly violates conformal invariance; however, we expect that
for any ordinary QFT, a conformal phase should exist for very high temperature.
Following \cite{Sasai}, for $\beta\ll 1$ we can consider \cite{DV, Gradsteyn}
\begin{eqnarray}
\sum_{\omega=1}^{\infty}\frac{\beta\omega}{e^{\beta\omega}-1}
&\rightarrow& \int_{0}^{\infty}\frac{dx}{\beta}\frac{x}{e^{x}-1} = \frac{\pi^{2}}{6\beta},
\\
\sum_{\omega=1}^{\infty}\ln\left(1-e^{-\beta\omega}\right)
&\rightarrow& \int_{0}^{\infty}\frac{dx}{\beta}\ln\left(1-e^{-x}\right) = - \frac{\pi^{2}}{6\beta},
\end{eqnarray}
and (\ref{Ss}) becomes
\begin{equation}
S(\beta \ll 1)= \frac{\pi^{2}}{3\beta}=2\pi\sqrt{\frac{\bar{N}}{6}},
\label{sbeta1}
\end{equation}
where we have defined
\begin{equation}
\bar{N}=\sum_{\omega}\omega\bar{N}_{\omega}=-\frac{\partial \ln Z}{\partial \beta},
\end{equation}
for $\bar{N}_{\omega}$ given in (\ref{bose}). The expression (\ref{sbeta1}) is
compatible with the Cardy formula for a central charge $c=1$.
This reflects the fact that the conformal phase of the system is recovered in this
limit and the value of the central charge corresponds to the single effective degree
of freedom of a $D-1$-brane on which the string ends.

Once one has the general structure presented earlier, it is possible to obtain the
free propagators for the model. Considering a fixed $\rho$ and following the TFD
procedure, the propagators can be arranged as a matrix
\begin{equation}
{\mathbb D}(t-t', \rho;\gamma) =
\left(
\begin{matrix}
\langle 0(\gamma)|
T[X(t,\rho)X(t',\rho)]| 0(\gamma)\rangle
&&&
\langle 0(\gamma)|
T[X(t,\rho)\widetilde{X}(t',\rho)]| 0(\gamma)\rangle
\\
\\
\langle 0(\gamma)|
T[\widetilde{X}(t,\rho)X(t',\rho)]| 0(\gamma)\rangle
&&&
\langle 0 (\gamma)|
T[\widetilde{X}(t,\rho)\widetilde{X}(t',\rho)]| 0(\gamma)\rangle
\end{matrix}
\right) ,
\end{equation}
where $T$ stands for the time ordering of the worldsheet operator products
in such a way that, as usual,
\begin{equation}
T[X(t,\rho)X(t',\rho)]= \Theta(t-t') X(t,\rho)X(t',\rho) + \Theta(t'-t) X(t',\rho)X(t,\rho)
\end{equation}
where $\Theta(t-t')$ is the Heaviside step function. As before, $X(t,\rho)$
and $\widetilde{X}(t,\rho)$ denote the solutions for the L and R
quadrants, respectively. Using (\ref{sol1}) and (\ref{soll}), the matrix
entries are written in terms of the following propagators
\begin{eqnarray}
D_{11}(t-t',\rho ;\gamma)&=&
\langle 0(\gamma)|X(t,\rho)X(t',\rho)| 0(\gamma)\rangle
\nonumber
\\
&=&\sum_{\omega > 0} \frac{1}{1-f_{\omega}}
\left\{
u_{\omega}(t,\rho)u^{*}_{\omega}(t',\rho)
+ f_{\omega}\, u^{*}_{\omega}(t,\rho)u_{\omega}(t',\rho)
\right\},
\\
D_{12}(t-t',\rho ;\gamma) &=&
\langle 0(\gamma)| X(t,\rho)\widetilde{X}(t',\rho)| 0(\gamma)\rangle
\nonumber
\\
&=&-\!\!\sum_{\omega > 0}\frac{1}{1-f_{\omega}}
\left\{
e^{2s_\omega} f^{\alpha}_{\omega}\, u_{\omega}(t,\rho)\tilde{u}_{\omega}(t',\rho)
+ e^{-2s_\omega} f^{\alpha^{*}}_{\omega} u^{*}_{\omega}(t,\rho)\tilde{u}^{*}_{\omega}(t',\rho)
\right\}\! ,
\\
D_{21}(t-t',\rho ;\gamma) &=&
\langle 0(\gamma)|\widetilde{X}(t,\rho)X(t',\rho)| 0(\gamma)\rangle
\nonumber
\\
&=& - \!\! \sum_{\omega > 0}\frac{1}{1-f_{\omega}}
\left\{
e^{2s_\omega} f^{\alpha}_{\omega}\,\tilde{u}_{\omega}(t,\rho) u_{\omega}(t',\rho)
+ e^{-2s_\omega} f^{\alpha^{*}}_{\omega} \tilde{u}^{*}_{\omega}(t,\rho) u^{*}_{\omega}(t',\rho)
\right\}\! ,
\\
D_{22}(t-t',\rho ;\gamma) &=&
\langle 0 (\gamma)|\widetilde{X}(t,\rho)\widetilde{X}(t',\rho)| 0(\gamma)\rangle
\nonumber
\\
&=&\sum_{\omega > 0} \frac{1}{1-f_{\omega}}
\left\{
\tilde{u}_{\omega}(t,\rho)\tilde{u}^{*}_{\omega}(t',\rho)
+ f_{\omega}\, \tilde{u}^{*}_{\omega}(t,\rho)\tilde{u}_{\omega}(t',\rho)
\right\}.
\end{eqnarray}
Considering the worldsheet defined in the boundary $\rho \rightarrow \rho_{c}$, the solutions (\ref{sol1}) and
(\ref{soll}) become
\begin{eqnarray}
X(t,\rho_{c})= \sum_{\omega > 0}
\sqrt{\frac{2\alpha'\beta}{\ell^{2}\omega\log(1/\epsilon)}}
\left[\frac{1-i\nu}{1-i\rho_{c}\nu}\left(\frac{\rho_{c}-1}{\rho_{c}+1}\right)^{\frac{i\nu}{2}}
e^{-i\omega t} a_{\omega}
\right.
\nonumber
\\
\left.
+ \frac{1+i\nu}{1+i\rho_{c}\nu}\!\left(\frac{\rho_{c}-1}{\rho_{c}+1}\right)^{\frac{-i\nu}{2}}
e^{i\omega t}a_{\omega}^{\dagger}\right],
\end{eqnarray}
for the R quadrant and
\begin{eqnarray}
\widetilde{X}(t,\rho_{c}) = \sum_{\omega > 0}
\sqrt{\frac{2\alpha'\beta}{\ell^{2}\omega\log(1/\epsilon)}}
\left[\frac{1+i\nu}{1+i\rho_{c}\nu}\left(\frac{\rho_{c}-1}{\rho_{c}+1}\right)^{\frac{-i\nu}{2}}
e^{i\omega t}\tilde{a}_{\omega}
\right.
\nonumber
\\
\left.
+ \frac{1-i\nu}{1-i\rho_{c}\nu}\left(\frac{\rho_{c}-1}{\rho_{c}+1}\right)^{\frac{i\nu}{2}}
e^{-i\omega t}\tilde{a}_{\omega}^{\dagger}\right],
\end{eqnarray}
for the L one. Therefore
\begin{eqnarray}
u_{\omega}(t,\rho_{c})=\tilde{u}_{\omega}^{*}(t,\rho_{c})
=\frac{1-i\nu}{1-i\rho_{c}}
\left(\frac{\rho_{c}-1}{\rho_{c}+1}\right)^{i\frac{\nu}{2}}
e^{-i\omega t},
\\
u^{*}_{\omega}(t,\rho_{c})=\tilde{u}_{\omega}(t,\rho_{c})
=\frac{1+i\nu}{1+i\rho_{c}}
\left(\frac{\rho_{c}-1}{\rho_{c}+1}\right)^{-i\frac{\nu}{2}}
e^{i\omega t},
\end{eqnarray}
Replacing the expressions written above in the propagators, we have
\begin{eqnarray}
D_{11}(t-t',\rho_{c};\gamma)&=&
\frac{2\alpha'\beta}{\ell^{2}\log( 1/ \epsilon)} \frac{1+\nu^{2}}{1+\rho^{2}_{c}\nu^{2}}
\sum_{\omega > 0} \frac{1}{\omega} \left\{ e^{-i\omega (t-t')}
+ 2\frac{f_{\omega}}{1-f_{\omega}} \cos\left(\omega(t-t')\right)
\right\}
\label{D11}
\\
D_{12}(t-t',\rho_{c};\gamma)&=&
-\frac{2\alpha'\beta}{\ell^{2}\log( 1/ \epsilon)}\frac{1+\nu^{2}}{1+\rho^{2}_{c}\nu^{2}}
\nonumber
\\
& &\times \sum_{\omega > 0} \frac{1}{\omega} \frac{1}{1-f_{\omega}}
\left\{e^{2s_{\omega}}f^{\alpha}_{\omega} e^{-i\omega (t-t')}
+ e^{-2s_{\omega}}f^{\alpha^{*}}_{\omega}e^{i\omega (t-t')}
\right\}
\label{D12}
\\
D_{21}(t-t',\rho_{c};\gamma)&=&
-\frac{2\alpha'\beta}{\ell^{2}\log( 1/ \epsilon)}\frac{1+\nu^{2}}{1+\rho^{2}_{c}\nu^{2}}
\nonumber
\\
& &\times \sum_{\omega > 0} \frac{1}{\omega} \frac{1}{1-f_{\omega}}
\left\{e^{2s_{\omega}}f^{\alpha}_{\omega} e^{i\omega (t-t')}
+ e^{-2s_{\omega}}f^{\alpha^{*}}_{\omega}e^{-i\omega (t-t')}
\right\}
\label{D21}
\\
D_{22}(t-t'\rho_{c};\gamma)&=&
\frac{2\alpha'\beta}{\ell^{2}\log( 1/ \epsilon)} \frac{1+\nu^{2}}{1+\rho^{2}_{c}\nu^{2}}
\sum_{\omega > 0} \frac{1}{\omega} \left\{ e^{i\omega (t-t')}
+ 2\frac{f_{\omega}}{1-f_{\omega}} \cos\left(\omega(t-t')\right)
\right\}
\label{D_22}
\end{eqnarray}
The expression (\ref{D11}) is the propagator commonly used when thermal
equilibrium is given. For example, making $f_{\omega} = e^{-\beta\omega}$
and $t'=0$, we have (\ref{D11}) in a perfect match with
the worldsheet thermal two point functions derived in \cite{deboer}
and considered in the calculation of the displacement of string endpoint,
establishing the connection with the standard Brownian motion.

\section{Concluding Remarks}

In this work we have studied the string propagating in an AdS Schwarzschild spacetime
from the point of view of a thermal theory in the worldsheet. In particular, the
approach to the Brownian motion of a quark in a CFT fluid at finite temperature
developed in Ref. \cite{deboer} was reinforced in this work since the main computations
were reproduced. The construction presented here is based on Israel-Maldacena's picture,
where the AdS-Schwarzschild spacetime is maximally extended along with the fundamental
string solution: we have extended it through natural gluing conditions imposed on an
effective surface near the event horizon. We have pointed out that the gluing conditions
of the string at the horizon define a boundary state, which is exactly the thermal vacuum
of the Thermo Field Dynamics (TFD), and the connection of this point of view with unitary
SU(1,1) TFD formulation  was  emphasized here. It was shown that all information about the
bulk and the board are present in the SU(1,1) parameters. As the string boundary state is
an entangled state, we calculated the entanglement entropy, which in this case coincides
with the thermodynamic entropy. We show that in, the high temperature limit, the result
agrees with the Cardy formula, reflecting the fact that the conformal phase is recovered
at high temperatures. Also, the canonical approach of TFD was explored to calculate the
worldsheet real time thermal Green's functions. An important aspect  of the framework
explored here is the presence of a string boundary state. This kind of state can be used
to study the  precise microscopic structure of the stretched horizon. In fact, as noticed in Ref. \cite{brane-closed-pure}, this state approaches the following pure
but coherent one:
\begin{equation}
\left|0(\gamma)\right\rangle \, \sim \,e^{\sum_{\omega=1}
a^{\dagger}_{\omega}\tilde{a}^{\dagger}_{\omega}}\left.
\left|0\right\rangle\!\right\rangle,
\end{equation}
in the limit $K \to 0$ (or $\alpha\to \infty$) of (\ref{entropiavacum}), so it describes a macroscopic
(semiclassical) object. Because the open string ends on this surface, we can
go to the closed string channel and this state can clearly be identified with
the state of a $D_p$-brane ($p=D-1$) \cite{braneopen}, which reinforces the idea
that the stretched horizon may be described as a real $D_p$-brane.
 A similar scenario was studied in Ref. \cite{MN} in the context
of  pp-wave time-dependent background, where it  was shown that, for asymptotically
flat observers, the closed string vacuum close to the singularity appears as a boundary state, which is in fact a D-brane described in the closed string channel.

In a forthcoming work we shall try to describe the Brownian motion and the stochastic
effects (ruled out by a Langevin equation) in terms of microscopic aspects of the string
gluing, and by considering interaction between both strings in the contact (horizon) surface.
In fact, the velocity of the string endpoint is high if it does not fall into the black hole,
so more than the quadratic order should be taken into account in the Nambu-Goto action as
the parameter $\epsilon (\equiv \rho-1)$ approaches to zero. A sketch to argument dissipative
effects as $\epsilon\to0$ in this framework is indeed the following: non-gaussian terms in
the action are not invariant under Bogoliubov transformations in general; so for
generic frames (accelerated with respect to the horizon), products of tilde with non-tilde
fields are induced by the Bogoliubov transformation of these terms, which typically describes
dissipation in TFD, \cite{netfd1,netfd2,mizutani, rindlerstretched, NV12}. A similar idea was applied to construct a string vertex state for a Rindler horizon \cite{rindlerstretched}.

We conclude this work by pointing out that some more refined interpretations arise from the present construction. For instance, it suggests that the dual picture of the AdS stretched horizon
might be described as an entangled state of two (decoupled) heavy quarks in the hydrodynamic
regime of CFT's fluids.

\begin{acknowledgments}
Daniel Luiz Nedel would like to thank CNPq, grant 501317/2009-0, for financial support. MBC was partially supported by: CONICET PIP 2010-0396 and ANPCyT PICT 2007-0849.
\end{acknowledgments}

\end{document}